# EXPERIMENTAL AND NUMERICAL EVALUATIONS OF THE VISCOELASTIC MECHANICAL BEHAVIOR OF ARTERIAL TISSUE

Xiaochang Leng[1], Xiaomin Deng[2], Suraj Ravindran[2], Addis Kidane[2], Susan M. Lessner[3], Michael A. Sutton[2], Tarek Shazly[2,4]

[1] School of Civil Engineering and Architecture, Nanchang University,
Jiangxi, 330031, People's Republic of China

[2] College of Engineering and Computing, Department of Mechanical Engineering
University of South Carolina,
Columbia, SC 29208, USA

[3] School of Medicine, Department of Cell Biology & Anatomy
University of South Carolina
Columbia, SC 29208, USA

[4] College of Engineering and Computing, Biomedical Engineering Program
University of South Carolina
Columbia, SC 29208

Running title: Viscoelastic mechanical behavior of arterial tissue
Corresponding author:
Xiaochang Leng, Ph.D.
Email: lengxc1984@163.com
Address: 999 University Avenue, Honggutan District, Nanchang,
Jiangxi, 330031, People's Republic of China

**ABSTRACT**

The viscoelastic properties of arterial tissue dictate vessel behavior in certain disease states, injury modalities, and during some endovascular procedures. In this study, we characterized the viscoelastic mechanical response of porcine abdominal aortic tissue via uniaxial mechanical experiments on vascular tissue samples coupled with full-field surface strain measurements using the 2-D digital image correlation (DIC) technique. The measured stress relaxation response and intimal surface strain field were used to identify material parameters for a proposed viscoelastic anisotropic (VA) constitutive model of the passive arterial wall. The obtained results show that the VA constitutive model is able to capture the viscoelastic mechanical behavior of the arterial tissue over clinically-relevant time scales. The identified material model and numerical simulations provide a comprehensive description of the passive viscoelastic tissue properties of the arterial wall, and a quantitative understanding of the spatial and directional variability underlying arterial tissue mechanical behavior.

# 1. Introduction

Soft biological tissues exhibit a range of mechanical behaviors under various loading conditions. Although not prominent under most physiological scenarios, most soft tissues can exhibit a viscoelastic response [1, 2]. For example, the dramatic, large deformation of arterial tissues under trauma or endovascular intervention (stenting or balloon angioplasty) elicits a viscoelastic response which manifests as a sensitivity to loading rate [3]. The inherent viscoelasticity of soft tissue, which affects the instantaneous maximum stress, enables it to reduce or eliminate damage during such deformations. Given the complex interplay between injury, interventional modalities, and the biomechanical behavior of impacted vessels, there remains a need for comprehensive characterization of the viscoelastic properties of arterial tissue.

The viscoelastic properties of arterial tissues have been considered in modeling the mechanical behavior of vulnerable arteries using the finite element method [4]. It has been shown that the viscoelasticity of arterial tissues plays an important role in the arterial response to vasoactive drugs vascular trauma [2, 5]. Quantifying the passive viscoelastic mechanical properties of arterial tissue can therefore elucidate the primary determinants of arterial tissue mechanical behavior under clinically-relevant conditions that differ from the normal physiological state, can provide guidance for evaluation of vasoactive therapies, and aid in the design of appropriate protocols for endovascular interventions.

Viscoelasticity of biological soft tissues results from the friction between collagen fibers and other matrix components (e.g., smooth muscle cells and elastin); the friction from the movement and permeation of water molecules between these matrix components and fibers [6]; and the viscoelastic response of isolated collagen fibrils [7]. A structure-motivated viscoelastic constitutive model of arterial tissues can characterize the local mechanical environment of resident vascular cells and provide insight into the mechanical implications of both vascular injury and clinical intervention.

Although there have been many investigations devoted to the elastic mechanical response and properties of arterial tissue [8-10], relatively few studies have focused on the viscoelastic response. A viscoelastic algorithm with the generalized Maxwell model, containing a spring and an arbitrary number of dashpot elements arranged in parallel, was incorporated into a hyperelastic model to predict the hysteresis of arterial tissues [11-13]. A viscoelastic model was proposed to simulate the 'stretch inversion phenomenon' in the low pressure regime and to illustrate the anisotropic viscoelastic characteristics of a fiber-reinforced composite material [14], yielding a viscoelastic anisotropic (VA) model [13, 15]. A transversely isotropic and hyperelastic material model, combined with a Kelvin-Voigt linear viscous model, was proposed to characterize the viscoelastic mechanical responses of native and fabricated biological tissues [16, 17]. Moreover, at the micro-scale level, the generalized Maxwell model was used to characterize the viscoelastic behaviors of a three-dimensional biopolymer network relevant to arterial tissue [18].

A uniaxial tensile experiment is often conducted to assess the mechanical response of vascular tissue, and the experimental results provide preliminary/comparative information on the effects of certain medical treatments or disease states [19-22]. Additionally, the digital image correlation (DIC) method, an optical metrological technique that is capable of measuring the full-field surface strain by comparing the gray level intensity values on a specimen's surface at two deformed states, has been applied for strain measurements of soft tissues [23]. Given the mechanical heterogeneity of vascular tissue under large deformation, it is advantageous to use DIC to capture the local strain field as tissue samples undergo deformation in order to assess more accurately the mechanical behavior.

In the current study, an integrated theoretical-experimental approach was taken to quantify the passive viscoelastic properties of porcine abdominal aorta. First, a viscoelastic anisotropic constitutive formulation based on extension of an existing constitutive model for arterial tissue was proposed. Then, experiments were carried out on porcine abdominal aortic specimens to measure the viscoelastic response. Finally, based on the experimental data, the values of the material parameters in the proposed viscoelastic anisotropic constitutive model were determined. These parameter values provide a quantitative understanding of the variations of the local viscoelastic properties of arterial tissue, including spatial heterogeneity and specimen orientation dependence.

## 2. Theoretical framework

### 2.1 A viscoelastic anisotropic model for arterial wall tissues

A viscoelastic anisotropic constitutive model for arterial tissue is proposed based on the idea of decomposition of the stress response to mechanical loading into three different parts: the volumetric response, the isochoric response (the response in an equilibrium state), and the viscoelastic response (the response in a non-equilibrium state) [24]. The stress response in an equilibrium state follows that of the Holzapfel-Gasser-Ogden (HGO) model [10], which is an hyperelastic anisotropic model for arterial tissue, and the stress response in a non-equilibrium state is incorporated with implementation of the general Maxwell model [15, 25].

#### 2.1.1 Hyperelastic anisotropic model

In the HGO hyperelastic anisotropic model, the arterial wall is considered to contain two families of collagen fibers. The free energy potential $\Psi(\boldsymbol{C}, \boldsymbol{H}_1, \boldsymbol{H}_2)$, which defined per unit reference volume of arterial tissue in a decoupled form, is expressed as [15]:

$$\Psi(\boldsymbol{C}, \boldsymbol{H}_1, \boldsymbol{H}_2) = \Psi_{vol}(J) + \overline{\Psi}(\overline{\boldsymbol{C}}, \boldsymbol{H}_1, \boldsymbol{H}_2) \qquad (1)$$

where $\boldsymbol{C}$ is the right Cauchy-Green strain tensor and $\overline{\boldsymbol{C}}$ denotes a modified counterpart, $\overline{\boldsymbol{C}} = \overline{\boldsymbol{F}^T}\overline{\boldsymbol{F}}$; $\overline{\boldsymbol{F}} = J^{-1/3}\boldsymbol{F}$, where $\boldsymbol{F}$ is the deformation gradient tensor, and $J = \det(\boldsymbol{F})$. The two structure tensors which depend on the fiber family direction vectors $\boldsymbol{a}_{01}$ and $\boldsymbol{a}_{02}$ are expressed as

$$H_1(a_{01}, \kappa) = \kappa I + (1 - 3\kappa)(a_{01} \otimes a_{01}) \quad (2)$$

$$H_2(a_{02}, \kappa) = \kappa I + (1 - 3\kappa)(a_{02} \otimes a_{02}) \quad (3)$$

$$[a_{01}] = \begin{bmatrix} \cos\gamma \\ \sin\gamma \\ 0 \end{bmatrix}, \quad [a_{02}] = \begin{bmatrix} \cos\gamma \\ -\sin\gamma \\ 0 \end{bmatrix} \quad (4)$$

where $I$ is the identity tensor; $\kappa$ is the dispersion parameter, describing the dispersion of the two families of collagen fibers in an arterial wall; $\kappa = 0$ when the two collagen fiber families are parallel to each other and $\kappa = 1/3$ when collagen fibers distribute isotropically; $\gamma$ denotes angle between the mean fiber orientation of one family of fibers and the circumferential direction of the aorta.

The volumetric part of the free potential energy, $\Psi_{vol}(J)$, is given by [26]

$$\Psi_{vol}(J) = \frac{1}{D}\left(\frac{J^2 - 1}{2} - \ln J\right) \quad (5)$$

where $\frac{1}{D}$ is analogous to the bulk modulus of the material.

The isotropic part of the free potential energy is given by

$$\bar{\Psi}(\bar{C}, H_1, H_2) = \bar{\Psi}_g(\bar{C}) + \bar{\Psi}_f(\bar{C}, H_1, H_2) \quad (6)$$

$$\bar{\Psi}_g(\bar{C}) = \frac{\mu}{2}(\bar{I}_1 - 3) \quad (7)$$

$$\bar{\Psi}_f(\bar{C}, H_1, H_2) = \frac{k_1}{2k_2}\left[e^{k_2[\kappa \bar{I}_1 + (1-3\kappa)\bar{I}_{41} - 1]^2} - 1\right] + \frac{k_1}{2k_2}\left[e^{k_2[\kappa \bar{I}_1 + (1-3\kappa)\bar{I}_{42} - 1]^2} - 1\right] \quad (8)$$

where $\bar{I}_1 = tr(\bar{C})$ denotes the first invariant of $\bar{C}$, and $\mu$ is a stress-like parameter, representing the shear modulus of the amorphous matrix; $\bar{I}_{41} = a_{01} \cdot \bar{C} a_{01}$ and $\bar{I}_{42} = a_{02} \cdot \bar{C} a_{02}$ are tensor invariants equal to the square of the stretch in the direction of $a_{01}$ and $a_{02}$, respectively. Note that the constitutive parameter $k_1$ is related to the relative stiffness of the fibers and $k_2$ is a dimensionless parameter.

**2.1.2 Viscoelastic anisotropic (VA) model**

The viscoelastic anisotropic (VA) model proposed in the current study extends the HGO model [10] by incorporating a viscoelastic component [14, 15]. Specifically, the VA model is composed of two distinct relaxation parts: the viscoelastic contributions from the ground matrix and from the collagen fibers, respectively.

In the VA model, the second Piola-Kirchhoff stress tensor $\boldsymbol{S}_{m+1}$ at time $t_{m+1}$ is given by [13, 18] (Fig.1):

$$\boldsymbol{S}_{m+1} = \left(\boldsymbol{S}_{vol}^{\infty} + \boldsymbol{S}_{g}^{\infty} + \boldsymbol{S}_{f}^{\infty} + \sum_{\alpha=1}^{n} \boldsymbol{Q}_{\alpha}\right)_{m+1} \tag{9}$$

where "∞" denotes the equilibrium condition when the time approaches infinity. The volumetric component (with subscript "*vol*") and the isochoric components (with subscript "*g*" for the ground matrix and subscript "*f*" for the collagen fibers) of the second Piola-Kirchhoff stress from the ground matrix and two families of collagen fibers have the form [13]:

$$\boldsymbol{S}_{vol}^{\infty} = \frac{1}{D}\left(J - \frac{1}{J}\right) J \boldsymbol{C}^{-1} \tag{10}$$

$$\boldsymbol{S}_{g}^{\infty} = 2\frac{\partial \bar{\Psi}_g(\bar{\boldsymbol{C}})}{\partial \boldsymbol{C}} = J^{-2/3}\mathbb{P}:\mu \boldsymbol{I} \tag{11}$$

$$\boldsymbol{S}_{f}^{\infty} = \sum_{i=1}^{2}\left[J^{-2/3}\mathbb{P}:\left(2k_1 e^{k_2 \bar{E}_i^2} \bar{E}_i H_i\right)\right] \tag{12}$$

where $\mathbb{P}$ is the projection tensor which has the form:

$$\mathbb{P} = \mathbb{I} - \frac{1}{3}\boldsymbol{C}^{-1}\otimes\boldsymbol{C} \tag{13}$$

where $\mathbb{I}$ is the fourth order identity tensor with the form in index notation [13]:

$$(\mathbb{I})_{IJKL} = \frac{\delta_{IK}\delta_{JL}+\delta_{IL}\delta_{JK}}{2} \tag{14}$$

$\delta_{IK}$ is the *Kronecker* delta. $\bar{E}_i$ ($i = 1, 2$) are the structure strain invariants and $\bar{\boldsymbol{h}}_i$ ($i = 1, 2$) are the modified structure tensors which can be expressed by

$$\bar{E}_i = \mathrm{tr}\bar{\boldsymbol{h}}_i - 1 \tag{15}$$

$$\bar{\boldsymbol{h}}_i = \kappa\bar{\boldsymbol{b}} + (1 - 3\kappa)(\bar{\boldsymbol{a}}_i\otimes\bar{\boldsymbol{a}}_i) \tag{16}$$

where $\bar{\boldsymbol{b}}$ denotes the modified left Cauchy-Green strain tensor, $\bar{\boldsymbol{b}} = \bar{\boldsymbol{F}}\bar{\boldsymbol{F}}^T$; $\bar{\boldsymbol{a}}_i = \bar{\boldsymbol{F}}\boldsymbol{a}_{0i}$ ($i = 1, 2$) represents the push-forward of $\boldsymbol{a}_{0i}$ through tensor $\bar{\boldsymbol{F}}$.

The non-equilibrium stress tensor $\boldsymbol{Q}_{\alpha\,(m+1)}$ is composed of components $\boldsymbol{Q}_{g\alpha\,(m+1)}$ from the ground matrix and $\boldsymbol{Q}_{f\alpha\,(m+1)}$ from the fibers at time $t_{m+1}$, which are expressed as

$$\boldsymbol{Q}_{g\alpha\,(m+1)} = \boldsymbol{\mathcal{H}}_{g\alpha\,(m)} + \beta_{g\alpha}\,exp\left(-\frac{\Delta t}{2T_{g\alpha}}\right)\left(\boldsymbol{S}_g^\infty\right)_{m+1} \tag{17}$$

$$\boldsymbol{Q}_{f\alpha\,(m+1)} = \boldsymbol{\mathcal{H}}_{f\alpha\,(m)} + \beta_{f\alpha}\,exp\left(-\frac{\Delta t}{2T_{f\alpha}}\right)\left(\boldsymbol{S}_f^\infty\right)_{m+1} \tag{18}$$

where the history term for the ground matrix and fibers are defined as:

$$\boldsymbol{\mathcal{H}}_{g\alpha\,(m)} = exp\left(-\frac{\Delta t}{2T_{g\alpha}}\right)\left[exp\left(-\frac{\Delta t}{2T_{g\alpha}}\right)\boldsymbol{Q}_{g\alpha\,(m)} - \beta_{g\alpha}\left(\boldsymbol{S}_g^\infty\right)_m\right] \tag{19}$$

$$\boldsymbol{\mathcal{H}}_{f\alpha\,(m)} = exp\left(-\frac{\Delta t}{2T_{f\alpha}}\right)\left[exp\left(-\frac{\Delta t}{2T_{f\alpha}}\right)\boldsymbol{Q}_{f\alpha\,(m)} - \beta_{f\alpha}\left(\boldsymbol{S}_f^\infty\right)_m\right] \tag{20}$$

The Cauchy stress tensor $\boldsymbol{\sigma}_{m+1}$ at time $t_{m+1}$ is

$$\boldsymbol{\sigma}_{m+1} = (J^{-1}\boldsymbol{F}\boldsymbol{S}\boldsymbol{F}^T)_{m+1} \tag{21}$$

In the above equations, $\Delta t$ is the time increment from time $t_m$ to $t_{m+1}$; $T_{g\alpha}$ ($\alpha = 1, 2, \ldots, n$ for the $n$ Maxwell elements) and $T_{f\alpha}$ are the relaxation times for the ground matrix and collagen fibers, respectively. Moreover, $\beta_{g\alpha}$ and $\beta_{f\alpha}$ are dimensionless parameters for the ground matrix and collagen fibers, respectively.

The viscous component $(\mathbb{C}_{vis})_{m+1}$ of the elasticity tensor for the ground matrix and collagen fibers at time $t_{m+1}$ are

$$(\mathbb{C}_{vis})_{g\,(m+1)} = (\mathbb{C}_g^\infty)_{m+1} \sum_{\alpha=1}^n \beta_{g\alpha} \exp\left(-\frac{\Delta t}{2T_{g\alpha}}\right) \tag{22}$$

$$(\mathbb{C}_{vis})_{f\,(m+1)} = (\mathbb{C}_f^\infty)_{m+1} \sum_{\alpha=1}^n \beta_{f\alpha} \exp\left(-\frac{\Delta t}{2T_{f\alpha}}\right) \tag{23}$$

And the total elastic tensor at time $t_{m+1}$ is given by

$$(\mathbb{C})_{m+1} = (\mathbb{C}_{vol}^\infty)_{m+1} + \left\{1 + \sum_{\alpha=1}^n \beta_{g\alpha} \exp\left(-\frac{\Delta t}{2T_{g\alpha}}\right)\right\}(\mathbb{C}_g^\infty)_{m+1} + \left\{1 + \sum_{\alpha=1}^n \beta_{f\alpha} \exp\left(-\frac{\Delta t}{2T_{f\alpha}}\right)\right\}(\mathbb{C}_f^\infty)_{m+1} \tag{24}$$

The spatial description of the elastic tensor at time $t_{m+1}$ is

$$\left(\mathbb{c}_{abcd}\right)_{m+1} = (J^{-1} F_{aA} F_{bB} F_{cC} F_{dD} \mathbb{C}_{ABCD})_{m+1} \tag{25}$$

## 3. Experiments

Experiments to characterize the viscoelastic response of arterial tissue were carried out with specimens obtained from porcine abdominal aorta as detailed below.

### 3.1 Vessel isolation

All tissue handling protocols were approved by the Institutional Animal Care and Use Committee at the University of South Carolina. Porcine abdominal aortas were obtained from a local slaughterhouse immediately after sacrifice of adult animals (8-12 months old, 75-125 lbs, male American Landrace Pigs) and placed in a phosphate buffer solution, cooled on ice, and transported immediately to the laboratory. Upon arrival, the abdominal aorta was isolated from the surrounding tissue, washed in phosphate buffered saline (PBS), and dissected free of perivascular tissue. A radial cut was made on the aorta in the axial direction, creating a flat sheet.

### 3.2. Specimen preparation and creation of surface speckle pattern

Six strips oriented at angles of 0° and 90° with respect to the circumferential vessel axis were cut from the two aortic samples using a surgical scalpel (Fig. 2). For the stress relaxation tests, the experimental sample set includes 5 strips of 0° orientation and 5 strips of 90° orientation. One strip each of 0°and 90° orientation were used for uniaxial tensile tests.

To apply the digital image correlation (DIC) method for surface deformation measurements, a speckle pattern was created on the specimens used for uniaxial tensile tests by intermittently spraying a tissue marking dye on the intimal surface of the specimen via an airbrush (Micron-B, IWATA) [27].

## 3.3 Mechanical experiments

### 3.3.1. Stress relaxation testing

Both ends of the tissue sample were attached on plastic plates with a tissue glue and then stabilized by the grips of the mechanical testing apparatus (Bose Enduratec 3200, Minnesota, MN) configured for stress relaxation testing.

To initiate mechanical testing, the sample was mechanically preconditioned via repeated quasi-static uniaxial tensile tests at a stretch ratio of 1.2. Samples were then stretched to multiple lengths (corresponding to stretch ratios of $\lambda_y$ = 1.2, 1.3 and 1.4) and held for 5 minutes to allow for stress relaxation, and then returned to the original length. During the loading-holding-unloading process, continuous data acquisition (scan rate of 50/sec) of the reaction force was made via the system load cell and software package (Wintest 4.0, Bose Electroforce, Minnesota, MN). The strips were intermittently hydrated by spraying with PBS prior to and during the stress relaxation process.

### 3.3.2 Uniaxial tensile testing

Specimen preparation and experimental setup are the same as for the stress relaxation tests (Section 3.3.1). Specimens were stretched (0.05 mm/s) until one of three axial stretch ratios (approximately 1.2, 1.3 and 1.4) was reached, and then immediately unloaded to the original length. This loading-unloading process was repeated several times for each specimen to generate several loading-unloading cycles. During each loading-unloading cycle, continuous data acquisition (scan rate of 50/sec) was made to record the reaction force via the system load cell and software package (Wintest 4.0, Bose Electroforce, Minnesota, MN).

For full-field surface strain measurements using the DIC method, a series of images of the specimen surface were recorded via a high-resolution camera (Point Grey Inc, Richmond, Canada) and a DIC system software package (Vic-Snap, Correlated Solutions Inc, Columbia, SC). An LED lamp was used to provide adequate lightening on the sample. Strain contours for each image were then determined through image analysis using the DIC system software (Vic-2D, Correlated Solutions Inc, Columbia, SC).

## 4. Numerical implementation

The VA model proposed in this study was implemented in the general-purpose finite element software ABAQUS [21] via user subroutines. One of the difficulties in the current study is identification of the values of material parameters in the VA model. These parameter values for porcine abdominal aortas are not available from the literature. Thus, a series of stress relaxation tests of porcine aorta specimens were performed and the material parameter values associated with the VA model were obtained through an automated inverse method implemented via a user-defined code written in MATLAB (MATLAB 2010, Mathworks Inc).

The material parameter values based on stress relaxation tests were identified via a least-squares fitting algorithm, performed using stress-relaxation test data for two specimen orientations. These material parameters were then used as initial guesses to identify the material parameter values based on uniaxial tensile tests, which was achieved through matching simulation predictions of the load vs. displacement curves for the first loading-unloading cycle. Using these identified parameter values, simulations of the subsequent loading-unloading cycles were performed to generate predictions would then be validated by the test data. The VA model was validated by comparing the simulation predictions with experimental measurements in terms of both the load-displacement curves and the surface strain field for various loading-unloading cycles.

## 5. Results

**5.1 Material parameter values based on stress relaxation tests**

Viscoelastic model parameter values were first estimated based on stress relaxation tests. This was done by finding a best-fit (Fig. 3) between experimental data and predictions of finite element analyses (FEA) for the variation of the reaction load with time in the stress relaxation tests (see Fig. 3 for two representative cases). A stress relaxation test includes three phases: loading, relaxation and unloading. Since the relaxation-phase response alone is not sufficient to characterize the viscoelastic anisotropic behavior of aortic tissues, test data for the entire loading-relaxation-unloading process was used to identify material model parameter values. The best-fit between test data and FEA predictions is achieved by minimizing the objective function $f$ [28] below:

$$f = \sum_{i=1}^{n} \left[\frac{F_{p_i} - F_{e_i}}{F_{e_i}}\right]^2 \qquad (30)$$

where $F_{p_i}$ and $F_{e_i}$ are, respectively, the reaction forces (loads) in response to stretching, as predicted by FEA or from experimental measurements, at the $i$-th observation time point. This minimization process was carried out through an automated inverse method implemented via a user-defined code written in MATLAB (MATLAB 2010, Mathworks Inc).

The same parameter value identification procedure can be used for other test data. For example, if surface strain data are used, the reaction forces in Eq. 30 will be replaced with surface strains, and if uniaxial tensile test data are used, then the load-stretch ratio response data (instead of the load-time variation data) will be used.

It is noted that, in the current study, it is found (see results below) that a single set of viscoelastic parameters $T_{g1}$, $\beta_{g1}$, $T_{f1}$ and $\beta_{f1}$, corresponding to $\alpha = 1$ in in Eq. 9 and one Maxwell element in Fig. 1, was sufficient to enable an adequate representation of the viscoelastic response observed in the experiments carried out in the current study. As such, identification of viscoelastic anisotropic model parameter values is shown in this paper for $\alpha = 1$ only.

All finite element analyses in the current study were carried out using the general-purpose finite element software ABAQUS [21], and the viscoelastic anisotropic model was implemented in ABAQUS via user subroutines. Implicit time integration was employed to model the loading, relaxation and unloading process. To ensure numerical accuracy, the time increment $\Delta t$ was changed according to the Newton-Raphson iteration criteria, and the stretching displacement loading rate was prescribed as 0.05 mm/s.

A verification example for finite element predictions based on the implemented VA material model is shown in the Appendix, in which numerical predictions were compared with analytcal solutions with satisfactory agreement.

The values of material model paramters associated with the viscoelastic anisotropic model that were identified through minimizing the objective function in Eq. 9 are listed in Table 1. It can be seen from Fig. 3 that the proposed viscoelastic anisotropic model, along with the set of identified paramter values in Table 1, was able to capture the viscoelastic response of the porcine abdominal aorta in both the circumferential and axial directions.

**5.2 Uniaxial tensile tests**

**5.2.1 Material parameter values based on uniaxial tensile tests**

Identification of material parameter values based on uniaxial tensile tests was carried out by using the same automated inverse analysis procedure as described in Section 5.1. Instead of a best-fit of the load-time variation data as in the case of stress relaxation tests, for the uniaxial tensile tests the objective was to match finite element simulation predictions of the load-stretch ratio curve with experimental measurements. In addition, the parameter values identified based on stress relaxation tests were used here as initial guess values to speed up the process.

It is noted that, although each uniaxial tensile test contained several loading-unloading cycles, only the load-displacement curve for the first cycle was used to identify the model parameter values [24]. Table 3 shows the material parameter values based on the first-cycle load-stretch ratio curve data for two specimens (with 0 and 90 degree orientations) from sample # 6 (thus the specimens were numbered #6-$0^o$ and #6-$90^o$). Comparisons of finite element simulation predictions with test data are shown in Fig. 4, which shows a good fit. The hysteresis phenomenon shown in Fig. 4 due to viscoelastic behavior is represented by the area between the loading and unloading phases.

**5.2.2 Validation based on uniaxial tensile tests**

Although Fig. 4 shows a good fit between simulation predictions and test data for the first loading-unloading cycle in uniaxial tensile tests, it does not represent an experimental validation. To provide a meaningful experimental validation for the simulation predictions based on the VA model and the identified model parameter values, the loading-unloading data after cycle 1 were utilized. Specifically, finite element simulations of the loading-unloading curves cycles 2 and 3 from the uniaxial tensile tests on specimens #6-$0^o$ and #6-$90^o$ were carried out. The VA model parameter values identified based on cycle 1 data were used as input to the finite element simulations. Comparisons of simulation predictions with test data are plotted in Fig. 5, which reveals an overall good agreement.

### 5.2.3 Validation based on surface strains data from DIC measurements

Validation of the VA model and its parameter values was also made by using surface strain data from DIC measurements. The normal strain data in the longitudinal direction from the uniaxial tensile tests, on specimen #6-0 (circumferential strip) and specimen #6-90 (axial strip), were chosen for the validation comparisons between finite element predictions and test data.

Figure 6 shows a direct comparison of surface strain contours. As can be seen, this figure does not lead to an obvious conclusion. This is because the strain field is basically uniform so the contours do not show a clear trend. A closer examination does reveal that the overall colors between simulation predictions and test data are consistent.

In order to gain a more quantitative comparison, the predicted and measured mean strain vs. stretch ratio for several loading and unloading cycles were plotted together in Fig. 7, which shows a good agreement between simulation predictions and test data.

## 6. Discussion

The aim of this study was to quantify the passive viscoelastic responses of abdominal aortic wall tissue and to validate a proposed viscoelastic anisotropic (VA) constitutive model to enable predictions of arterial wall stresses and strains under physiological or pathophysiological loading conditions. The passive viscoelastic responses were studied by considering the reaction force-relaxation time relationship in stress relaxation tests, the load-stretch ratio curves for loading-unloading cycles in uniaxial tensile tests, and surface strain fields of arterial tissues in uniaxial tensile tests at fixed global circumferential stretch ratios. A viscoelastic anisotropic (VA) constitutive model for arterial walls was proposed. The ability of the proposed model to capture viscoelastic behaviors of porcine abdominal aorta tissues observed from the stress relaxation tests and the uniaxial tensile tests was demonstrated.

The VA model parameter values were identified through an automated inverse analysis procedure by matching finite element simulation predictions with experimental measurements of the observed viscoelastic responses from the stress relaxation tests and the uniaxial tensile tests. The proposed VA model, along with the identified model parameter values, will enable the solutions of boundary-value problems in terms of predicting arterial mechanical responses and arterial wall stress distributions under various loading scenarios.

Specifically, the stress relaxation tests were performed on porcine abdominal aorta samples in the circumferential and axial directions, and the VA model parameter values for porcine abdominal aorta were identified through a least-square curve fitting algorithm. Three different stretch ratios were used for stress relaxation tests. The VA model parameter values obtained from the least-square curve fitting algorithm allow the VA model predictions to match well all measured stress vs. time curves for all three stretch ratios, demonstrating good agreement between VA model predictions and experimental data.

Some of the VA model parameters are adopted from a hyperelastic anisotropic model in the literature [10]. The values of these model parameters that were identified in this study are within the range of the published results from the literature [15]. In addition, the ratio of length to width of the tested specimens is more than 4 in order to eliminate the effects of boundary constraints [29, 30]. From previous studies, load decay occurs when the aortic tissue is stretched beyond a critical stretch ratio, which may affect the stress relaxation response [31-33]. Thus the maximum stretch ratio for the stress relaxation tests should less than 1.4 , in order to prevent damage in the arterial tissue for both circumferential and axial specimens [34].

The stress relaxation processes for each stretch ratio showed a rapid initial decay of stress, which was followed by a slow decrease in stress during the entire relaxation period of 300 s, until an equilibrium condition was reached. The magnitude of the stress decay during stress relaxation tests increased when the initial stress level was increased, which correlates to the numbers of recruited collagen fibers [35]. The properties of the dependence of stress relaxation on the initial stress level are in accordance with other studies [36]. The porcine aorta tissues were shown to have a short relaxation time in the range of 0.01~1 s and a long relaxation time in the range of 40~200 s, which are related to the viscoelastic behavior of the ground matrix and collagen fibers, respectively. The long relaxation time mechanism of collagen fibers is in agreement with the fact that bundles of fibers or individual fibers have shown to have slow and stable behavior under loading conditions [4, 37, 38]. Similar results were found that the time constant for native collagen fibrils during the assessment of viscoelastic properties is in the range of 60~250 s [39]. The material parameter values obtained in the current study from the stress relaxation tests are diversely distributed. This is believed to be attributable to variations in the microstructures of the specimens, as a result of variations in the properties of the matrix material, and in fibers density, distribution and configuration.

For the anisotropic mechanical response of aortic tissue, the specimens with axial orientation are more compliant than those with circumferential orientation [40]. This specimen-orientation dependence is consistent with the organization of the collagen fibers which sustain almost the entire mechanical response at high stretch ratio [2, 15, 41]. Moreover, the heterogeneity revealed from DIC strain measurements was obvious during the uniaxial tensile tests. At low stretch ratio, the normal strain distribution has a small variation near 0.2 and 0.27 for stretch ratios less than 1.23 and 1.21 of specimens from the circumferential direction and axial direction, respectively. This phenomenon is also consisitent with the distribution of the collagen fibril orientation and with the observation that the normal strain distribution had a large variation when the stretch ratio was increased and more fibers were recruited and started to take up load [18, 41].

Variations of local mechanical response due to spatial heterogeneity, specimen variability and orientation-dependence are clearly reflected in the identified material parameter values shown in Tables 1, 2 and 3. The current study provides a way of understanding and quantifying such variations by performing mechanical experiments and subsequent material parameter value identification analyses on specimens taken from different arterial vessel samples and/or from different regions and orientations of the same sample. In principle, statistical distributions of the material parameter values can be achieved using this approach.

There are several limitations that should be considered in the interpretation of the results. Firstly, the proposed VA model simplified the three-layer arterial wall as a one-layer wall and neglected variations in mechanical response through the wall thickness. Additional histological studies could provide insight into the microstructural configuration of the abdominal aortic wall and aid in construction of a layer-specific constitutive formulation for the VA model. Secondly, we only investigated the passive viscoelastic behavior of the arterial wall, neglecting the contribution of smooth muscle contraction. It has been shown that the basal tone of the vascular smooth muscle modulates the overall mechanical behavior of the arterial wall. Subsequent studies assessing the contribution of smooth muscle contractile states to various chemical and mechanical stimuli will elucidate the abdominal arterial wall mechanical response in conditions of health and disease.

## 7. Conclusions

To our knowledge, this study is the first to quantify the viscoelastic mechanical responses of porcine abdominal aorta. Viscoelastic passive mechanical properties were identified via uniaxial tensile testing coupled with the digital image correlation technique. Mechanical test data were used to demonstrate the utility of a proposed viscoelastic anisotropic (VA) constitutive model.

Comparisons of simulation predictions with experimental results in terms of the load vs. stretch ratio curve and intimal surface strain contours reveal that the VA constitutive model is able to capture the passive viscoelastic mechanical responses of arterial tissues. Quantifying the viscoelastic mechanical responses and validating the VA model are necessary to better understand aberrant mechanical performance of arterial tissue, facilitating predictions of stress and strain distribution of arterial tissues under various degrees of deformation and providing framework to design suitable surgical and endovascular interventions in this vital region of the circulatory system.

**Acknowledgments**

The authors gratefully acknowledge the sponsorship of NSF (award # CMMI-1200358) and partial support by a SPARC Graduate Research Grant from the Office of the Vice President for Research at the University of South Carolina.

**Appendix: Verification of viscoelastic anisotropic (VA) model**

The VA model is verified by comparing numerical and analytical results under the assumption that the material is under plane strain deformation. A unit-cubic model of fiber-reinforced material was implemented with two families of fibers distributed $47^0$ with respect to the circumferential direction (Fig.2a). The surfaces at the x-axis and z-axis are fixed along the x and z directions, respectively. The assumption of material incompressibility, implying that $\lambda_x \lambda_y \lambda_z = 1$. When the cubic model was stretched along the x-axis with a stretch ratio $\lambda_x$, the deformation gradient was calculated as follows:

$$\boldsymbol{F} = \begin{bmatrix} \lambda_x & 0 & 0 \\ 0 & 1/\lambda_x & 0 \\ 0 & 0 & 1 \end{bmatrix} \tag{A-1}$$

Substituting (9) into (21), the Cauchy stress was calculated,

$$\boldsymbol{\sigma}_{m+1} = \left(\boldsymbol{\sigma}^{\infty}_{vol} + \boldsymbol{\sigma}^{\infty}_{g} + + \boldsymbol{\sigma}^{\infty}_{f} + \boldsymbol{\sigma}_{visf1} + \boldsymbol{\sigma}_{visg1}\right)_{m+1} \tag{A-2}$$

For the volumetric component of the Cauchy stress tensor,

$$\boldsymbol{\sigma}^{\infty}_{vol} = \frac{1}{D}\left(J - \frac{1}{J}\right) J \boldsymbol{I} = -p \boldsymbol{I} \tag{A-3}$$

Since the surfaces perpendicular to the y-axis are traction free,

$$0 = \sigma_{m+1\,(yy)} = -p + \left(\sigma^{\infty}_{g} + + \sigma^{\infty}_{f} + \sigma_{visf1} + \sigma_{visg1}\right)_{m+1\,(yy)} \tag{A-4}$$

where $p = \left(\sigma^{\infty}_{g} + + \sigma^{\infty}_{f} + \sigma_{visf1} + \sigma_{visg1}\right)_{m+1\,(yy)}$.

With the material parameters of specimen #6-0 from Table 3, the comparison of Cauchy stresses in x direction between analytical and numerical results is shown in Fig. A1. The results show that the numerical predictions of stress vs. stretch ratio curves match the analytic outputs.

**Tables**

**Table 1**. Viscoelastic anisotropic model parameter values for porcine aortas, based on stress relaxation test data for 0 degree specimen orientation (loading along circumferential direction)

| Sample | $\mu$ [kPa] | $k_1$ [kPa] | $k_2$ | $\gamma$ [°] | $\kappa$ | $T_{g1}$ | $\beta_{g1}$ | $T_{f1}$ | $\beta_{f1}$ | Residual ($f$) |
|---|---|---|---|---|---|---|---|---|---|---|
| #1-0 | 10.89 | 139.33 | 0.114 | 41.81 | 0.26 | 0.02 | 3.47 | 98.11 | 0.22 | 0.042 |
| #2-0 | 25.62 | 126.76 | 0.03 | 33.17 | 0.25 | 0.02 | 3.47 | 92.65 | 0.38 | 0.025 |
| #3-0 | 2.64 | 25.38 | 0.003 | 22.72 | 0.10 | 0.059 | 10.35 | 138.82 | 0.16 | 0.013 |
| #4-0 | 2.81 | 46.88 | 0.001 | 28.89 | 0.18 | 0.02 | 0.003 | 62.25 | 0.18 | 0.008 |
| #5-0 | 18.54 | 48.13 | 0.01 | 24.67 | 0.18 | 0.02 | 0.003 | 71.95 | 0.27 | 0.003 |
| Mean | 12.1 | 77.3 | 0.032 | 30.25 | 0.19 | 0.028 | 3.46 | 92.76 | 0.24 | 0.018 |
| SD | 8.96 | 46.4 | 0.043 | 6.81 | 0.06 | 0.016 | 3.78 | 26.51 | 0.08 | 0.014 |

**Table 2**. Viscoelastic anisotropic model parameter values for porcine aortas, based on stress relaxation test data for 90 degree specimen orientation (loading along axial direction)

| Sample | $\mu$ [kPa] | $k_1$ [kPa] | $k_2$ | $\gamma$ [°] | $\kappa$ | $T_{g1}$ | $\beta_{g1}$ | $T_{f1}$ | $\beta_{f1}$ | Residual ($f$) |
|---|---|---|---|---|---|---|---|---|---|---|
| #1-90 | 10.2 | 46.7 | 0.16 | 52.58 | 0.11 | 15.41 | 0.34 | 30.86 | 0.27 | 0.002 |
| #2-90 | 6 | 65.1 | 0.001 | 63.39 | 0.26 | 100.45 | 0.25 | 68.97 | 0.13 | 0.004 |
| #3-90 | 18 | 62.7 | 0.01 | 61.44 | 0.27 | 0.10 | 22.48 | 94.41 | 0.23 | 0.003 |
| #4-90 | 26.6 | 141.7 | 10.41 | 43.37 | 0.28 | 1.49 | 2.62 | 99.80 | 0.24 | 0.002 |
| #5-90 | 7.5 | 46.7 | 0.33 | 55.95 | 0.20 | 0.10 | 0.01 | 113.56 | 0.19 | 0.003 |
| Mean | 13.7 | 72.6 | 2.18 | 55.35 | 0.22 | 23.51 | 5.14 | 81.52 | 0.21 | 0.003 |
| SD | 7.7 | 35.4 | 4.12 | 7.12 | 0.06 | 38.90 | 8.72 | 29.16 | 0.05 | 0.001 |

Table 3. Material parameters of VA model identified from specimens of sample #6.

| Sample | $\mu$ (kPa) | $k_1$ (kPa) | $k_2$ | $\gamma$ (angle) | $\kappa$ | $T_{g1}$(s) | $\beta_{g1}$ | $T_{f1}$(s) | $\beta_{f1}$ | Residual ($f$) |
|---|---|---|---|---|---|---|---|---|---|---|
| #6-0 | 65 | 565 | 1.2 | 51.7 | 0.2 | 0.05 | 4.28 | 25.7 | 1.2 | 0.088 |
| #6-90 | 35 | 6000 | 0.8 | 38.5 | 0.05 | 10 | 0.5 | 75 | 1 | 0.196 |

**Figure Legends**

**Figure 1.** Schematic model of a viscoelastic material.

**Figure 2.** Schematic of the experiments. (a) A radial cut was made on the porcine abdominal aorta, and strips oriented at 0° and 90° with respect to the circumferential vessel axis were obtained by cutting the aorta in different orientations; (b) Experimental setup of uniaxial tensile tests; (c) Specimen setup (zoomed-in view).

**Figure 3.** Comparison of experimental data and finite element best-fit predictions for representative stress relaxation test cases (porcine aorta samples group #2): (a) 0 degree specimen orientation (loading along circumferential direction) and (b) 90 degree specimen orientation (loading along axial direction).

**Figure 4.** Comparison of predicted and measured load-stretch ratio curve of the first loading-unloading cycle from uniaxial tensile tests: (a) specimen #6-0 (circumferential strip) (b) specimen #6-90 (axial strip).

**Figure 5.** Comparison of predicted and measured load-stretch ratio curves of 2nd and 3rd loading-unloading cycles from uniaxial tensile tests on specimen #6-0 (circumferential strip) and specimen #6-90 (axial strip).

**Figure 6.** Normal strain contours ($E_{yy}$, strain in the longitudinal direction) for specimens of group #6 (circumferential and axial strip) at three global stretch ratios from DIC measurements (a) and numerical predictions (b).

**Figure 7.** Plots of mean strain ($E_{yy}$) vs. stretch ratio, from simulation predictions (red line) and test data (blue line), for specimens #6-0 (circumferential strip) during the (a) 1$^{st}$, (c) 2$^{nd}$ and (e) 3$^{rd}$ cycles, and specimen #6-90 (axial strip) during the (b) 1$^{st}$, (d) 2$^{nd}$ and (f) 3$^{rd}$ cycles.

**Figure A.1.** Illustrative plots of Cauchy stress-stretch relationship in x-axis direction for analytical (empty circle) and numerical results (red line).

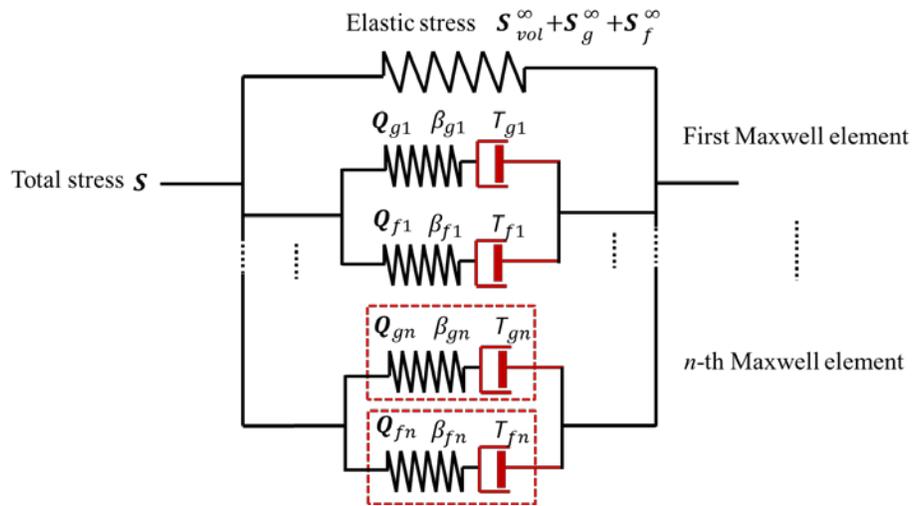

**Fig. 1.**

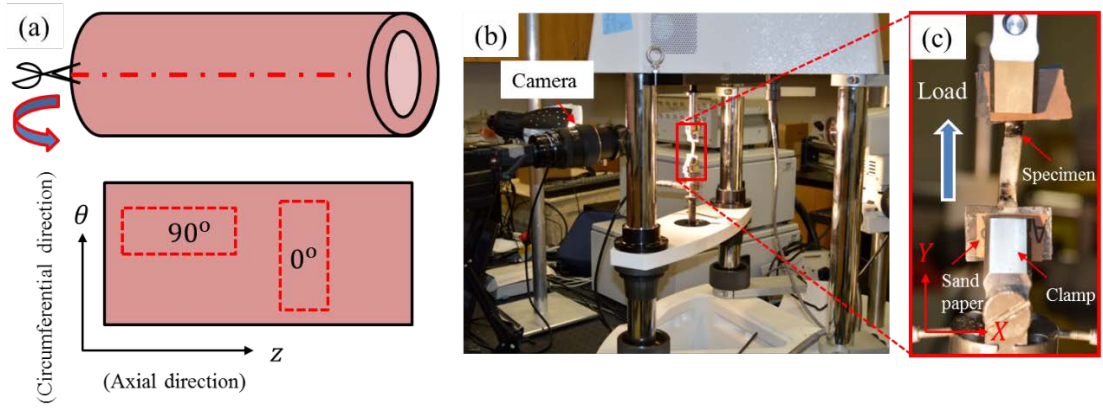

**Figure 2.**

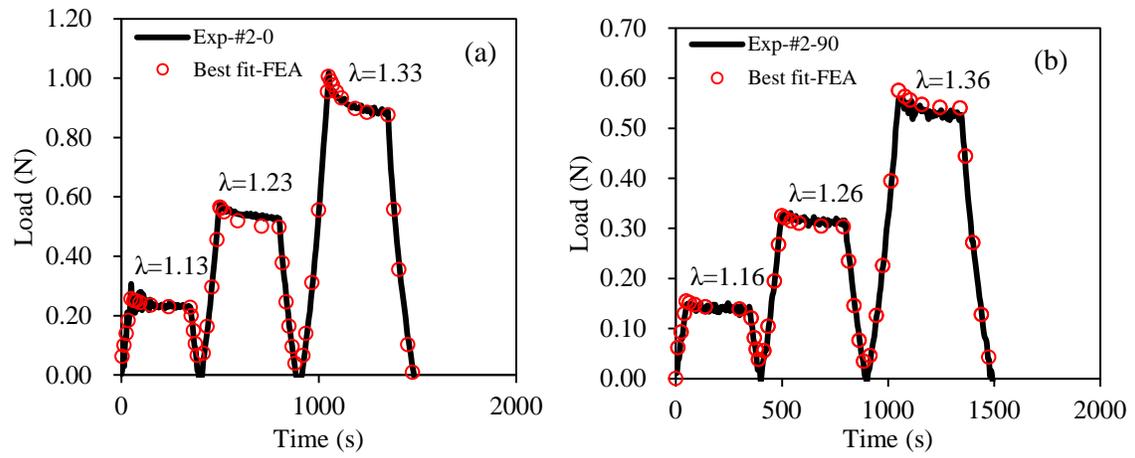

**Figure 3.**

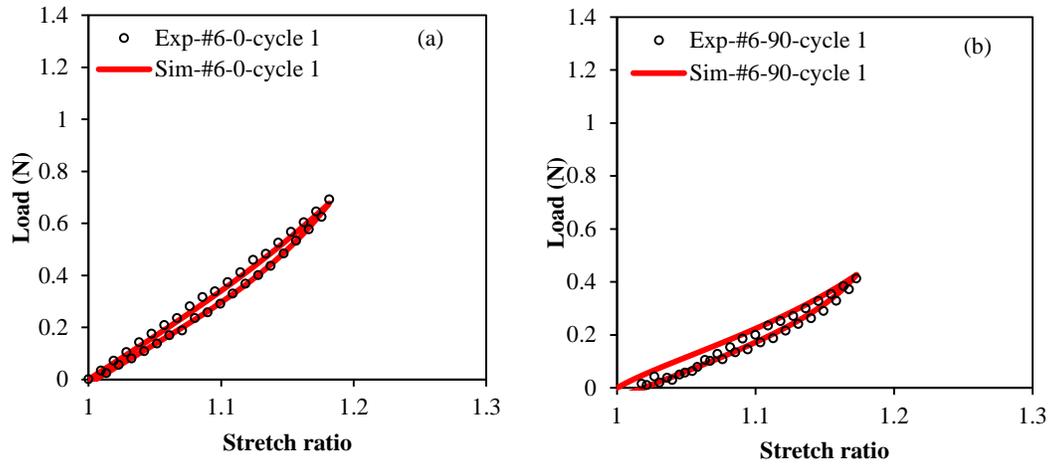

**Figure 4.**

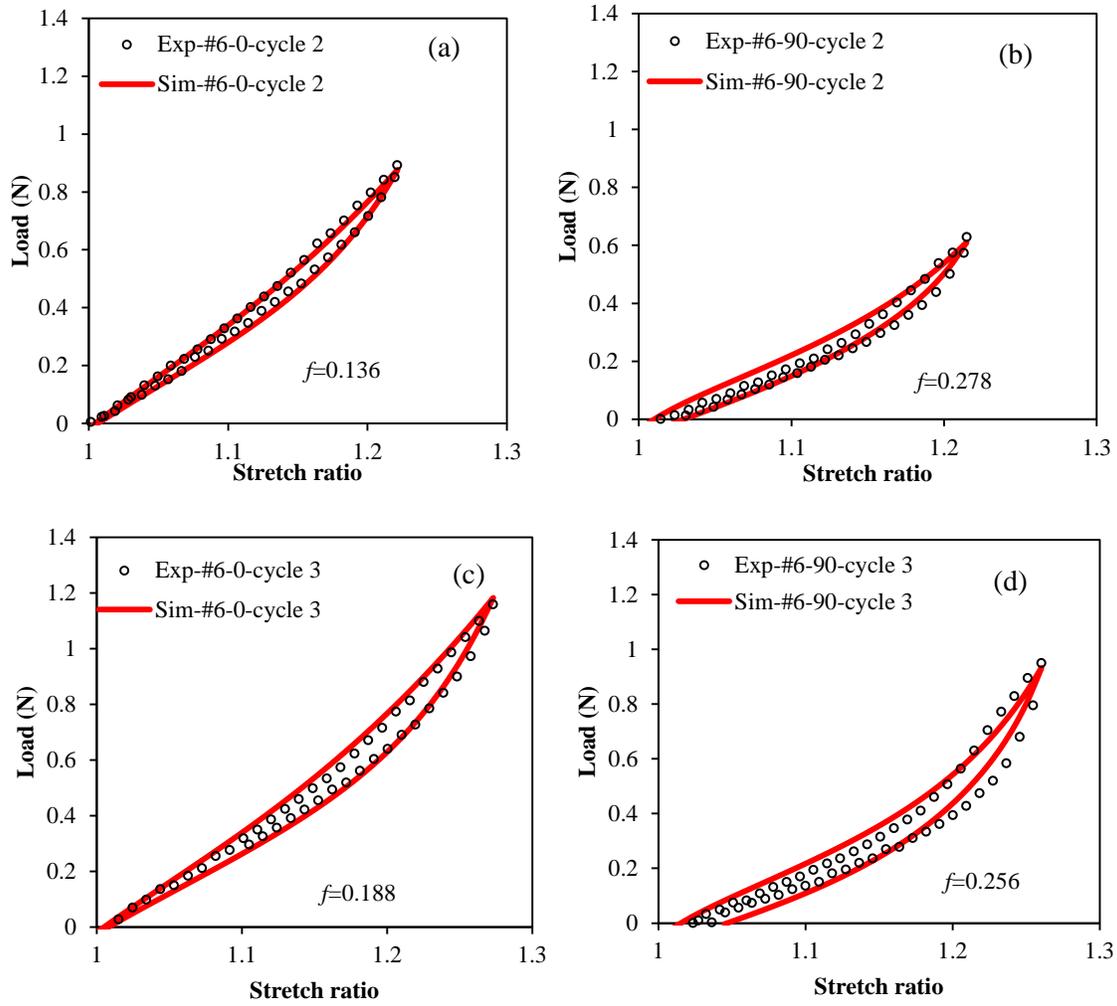

**Figure 5.**

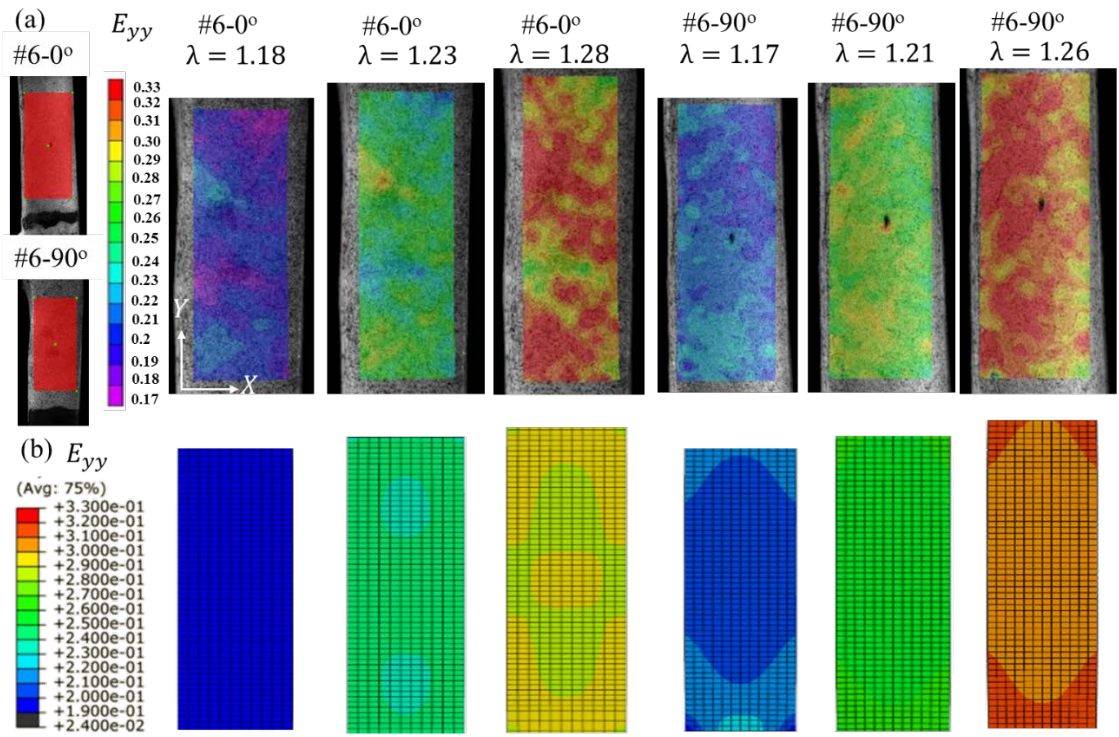

**Figure 6.**

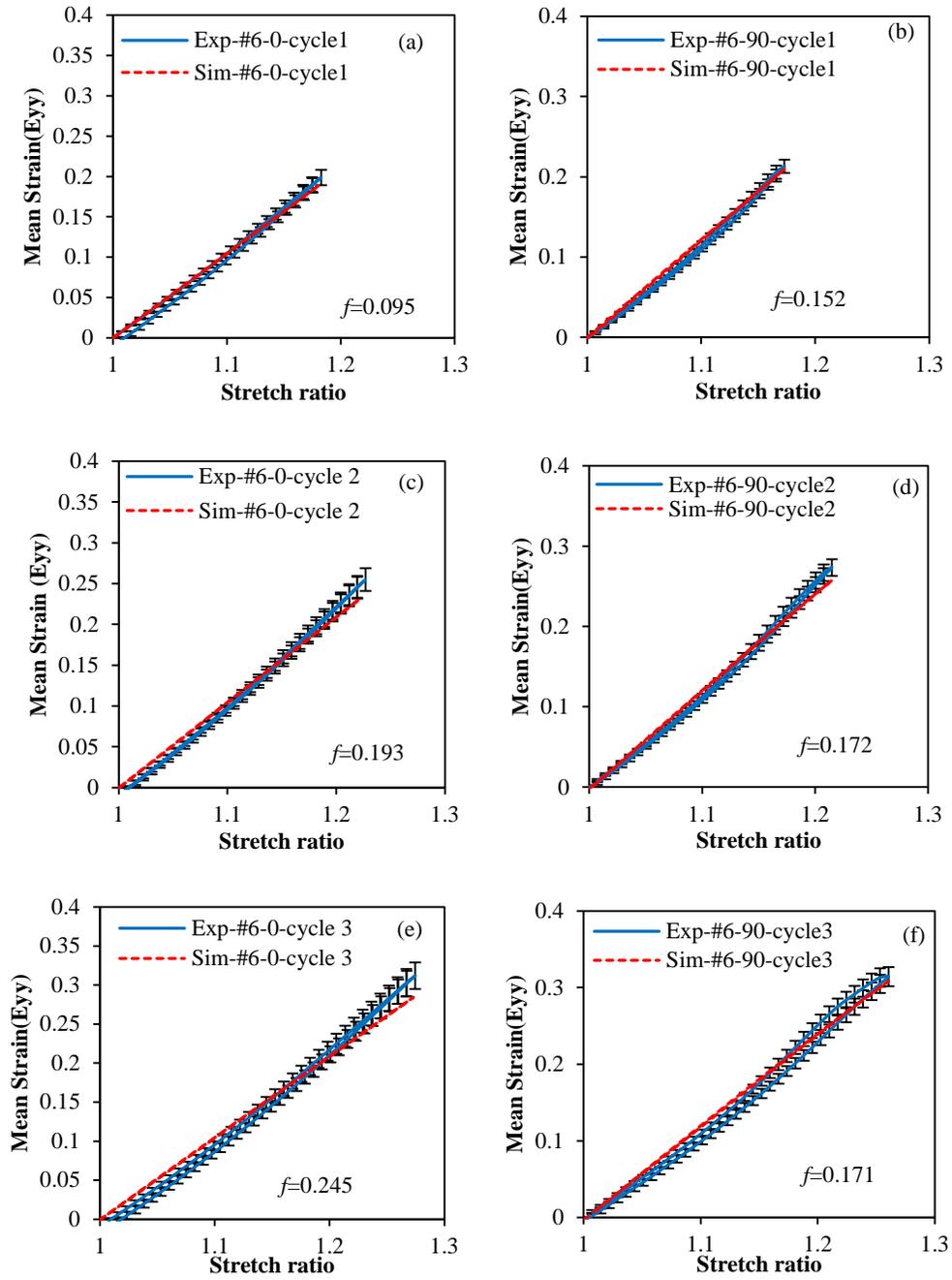

**Figure 7.**

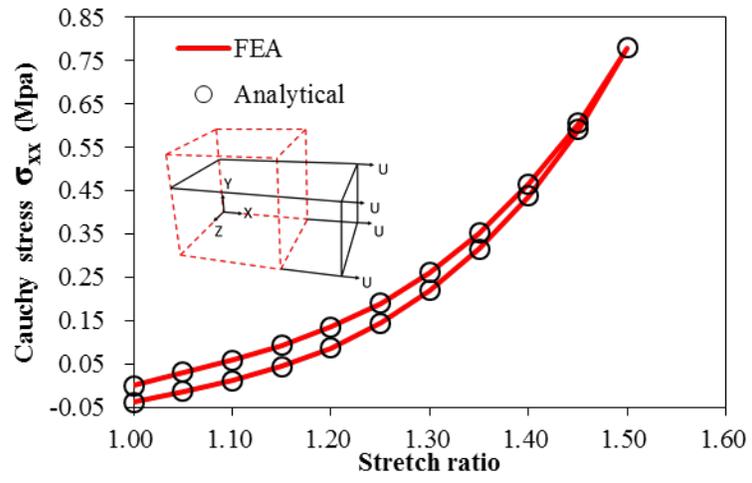

**Figure A.1.**